\documentclass[12pt]{revtex4}
\usepackage{amsmath}
\usepackage{graphicx}
\begin{document}

\title[Entropic transport of finite size particles]{Entropic transport of finite size particles}

\author{W Riefler}
\affiliation{$^{1}$Institut f\"ur Physik,
   Universit\"at Augsburg,
   Universit\"atsstr. 1,
   D-86135 Augsburg, Germany}
\author{G Schmid}
\affiliation{$^{1}$Institut f\"ur Physik,
   Universit\"at Augsburg,
   Universit\"atsstr. 1,
   D-86135 Augsburg, Germany}
\author{P S Burada}
\affiliation{Max-Planck Institute f\"ur Physik komplexer Systeme,
   N\"othnitzer Str. 38, 01187 Dresden, Germany}

\author{P H\"anggi}
\affiliation{$^{1}$Institut f\"ur Physik,
   Universit\"at Augsburg,
   Universit\"atsstr. 1,
   D-86135 Augsburg, Germany}

\begin{abstract}
Transport of spherical Brownian particles of finite size possessing radii $R\le R_\mathrm{max}$
through narrow channels with varying cross-section area is
considered. Applying the so-called Fick-Jacobs approximation, i.e.
assuming fast equilibration in orthogonal direction of the channel
axis, the 2D problem can be described by a 1D effective dynamics in
which bottlenecks cause entropic barriers. Geometrical confinements
result in entropic barriers which the particles have to overcome in
order to proceed in transport direction. The analytic findings for
the nonlinear mobility for the transport are compared with precise
numerical simulation results. The dependence of the nonlinear
mobility on the particle size exhibits a striking resonance-like
behavior as a function of the relative particle size
$\rho=R/R_\mathrm{max}$; this latter feature renders possible  new
effective particle separation scenarios.
\end{abstract}

\maketitle

\section{Introduction}

The diffusive behavior of Brownian particles depends mainly
on their size, the interaction between them, and the environment where they are situated in.
If, in addition to these characteristics, particles are confined within narrow, tortuous
structures such as nanopores, zeolites, biological
cells and microfluidic devices, the restriction of the space available
for the particles will cause entropic barriers that will have strong impact on the diffusive behavior
 (cf. Ref. \cite{chemphyschem} and references therein).
Effective
control schemes for transport in these systems require a detailed
understanding of the diffusive mechanisms involving small objects and,
in this regard, an operative measure to gauge the role of
fluctuations.
The study of these transport phenomena is, in many
respects, equivalent to an investigation of geometrical constrained
Brownian dynamics.
As the role of inertia for the motion of the particles through these
structures can typically be neglected, the Brownian dynamics can
safely be analyzed by solving the Smoluchowski equation in the domain
defined by the available free space upon imposing reflecting boundary
conditions at the domain walls.

However, solving the
boundary problem in the case of nontrivial, corrugated domains
presents a difficult task. A way to circumvent this difficulty
consists in coarsening the description by reducing the dimensionality
of the system considering only the main transport direction, but taking
into account the physically available space by means of an entropic
potential \cite{chemphyschem, reguera_prl,  biosystems}. The resulting
kinetic equation for the probability distribution, the so-called
Fick-Jacobs equation, is similar in form to the Smoluchowski equation,
but contains now entropic contributions leading to genuine dynamics
which distinctly differs from those observed for purely energetic
potentials.

The driven transport of particles across bottlenecks
\cite{chemphyschem,reguera_prl,biosystems,burada_pre}, such as ion
transport through artificial nanopores or artificial ion pumps
\cite{siwy2002,siwy2005, kosinska2008,dorp2009} or in biological channels
\cite{kullman2002,berezhkovskii2005,berezhkovskii2007,berezhkovskii2009}, are
striking examples where the diffusive transport is regulated by
entropic barriers.  In addition, geometrical confinements and entropic
barriers play also a prominent role in the context of the Stochastic
Resonance phenomenon \cite{gammaitoni, burada_prl, burada_epjb,
  burada_epl, ghosh2010}.

Our objective with this work is to investigate the mobility of
noninteracting spherical Brownian particles in channels with varying cross-section
width. In particular, we are interested in the influence of the
particle size on the transport within a periodic entropic potential
exhibiting barriers which arise from the geometrical restrictions.

The paper is organized as follows:  in section 2 we introduce the model and
define the theoretical and numerical problem. Further on, in section 3
we present the basic principles of the Fick-Jacobs approximation
allowing for reducing the two-dimensional problem to an
one-dimensional one. The results are presented in section 4. Finally,
we give the main conclusions in section 5.

\section{Modelling}

\begin{figure}[t]
  \centering
  \includegraphics{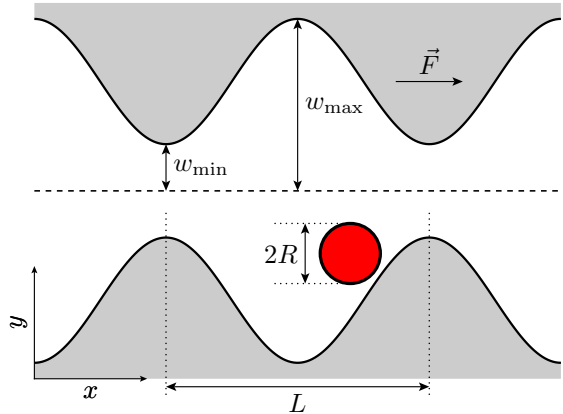}
  \caption{(Color online) Sketch of the 2D periodic channel with
    periodicity $L$, the minimum half channel width
    $\omega_{\mathrm{min}}$ and the maximum
    half channel width $\omega_{\mathrm{max}}$. The spherical Brownian particle
    of radius $R$ is subjected to the force $\vec{F}$.}
  \label{fig:tube}
\end{figure}

Transport through pores or channels (like the one depicted in
Fig. \ref{fig:tube}) may be caused by different particle concentrations
maintained at the ends of the channel, or by the application of
external forces acting on the particles. Here, we exclusively consider
the case of force driven transport of spherical particles of radius
$R$. The external force $\vec{F} = F \vec{e}_{x}$ is pointing parallel to the
direction of the channel axis. As small deviation from this assumption does not affect
our results, certainly not within the limits of validity of the Ficks-Jacob approximation.
Moreover, we shall assume low concentrations of spherical particles such that particle-particle
interactions and all hydrodynamic interaction effects can consistently be neglected.

\subsection{Dynamics inside the channel}

In general the dynamics of a suspended
Brownian particle is overdamped \cite{Purcell1977} and well described
by the Langevin equation:
\begin{equation}
  \label{eq:dle}
  \eta_{R} \frac{\mathrm{d} \vec{r} }{\mathrm{d} t } = \vec{F} +
  \sqrt{ \eta_{R} k_{\mathrm{B}}T} \vec{\xi}(t)\, ,
\end{equation}
where $\vec{r}$ denotes the center position of the spherical particle in the two-dimensional
channel, $k_{\mathrm{B}}$ the Boltzmann constant, $T$ the
temperature and $\vec{\xi}(t)$ is the  standard 2D Gaussian noise with $\langle
\vec{\xi}(t) \rangle = 0$ and $\langle \xi_{i}(t) \xi_{j}(t') \rangle
= 2 \delta_{i j} \delta(t -t')$ for $i,j = x,y$. The friction coefficient $\eta_{R}$ is given by
Stokes' law:
\begin{equation}
  \label{eq:friction}
  \eta_{R} = 6 \pi \nu R
\end{equation}
and depends on the shear viscosity $\nu$ of the fluid and the particle
radius $R$. In addition to Eq.~\ref{eq:dle} the full problem is set up
by imposing reflecting boundary conditions at the channel walls. The
boundary of the 2D periodic channel which is mirror symmetric
about its $x$-axis is given by the periodic function $y=\pm \omega(x)$ with
$\omega(x+L) = \omega(x)$ where $L$ is the periodicity of the
channel. $\omega_{\mathrm{min}}$ and $\omega_{\mathrm{max}}$ refer to
the half of the maximum and minimum channel width, respectively.

To further simplify the treatment of this problem, we introduce
dimensionless variables. We measure all lengths in units of the
periodicity of the channel, i.e. $x = x' L$. As a unit of time
$\tau$ we choose twice the time it takes the largest transporting
particle to diffusively cover the distance $L$ which is given by
$\tau = L^{2} \eta_{\max} / (k_{\mathrm{B} T})$, hence $t = \tau
t'$. The largest transportable particle is the particle with radius
$R_\mathrm{max}=\omega_\mathrm{min}$. Accordingly, the friction
coefficient of a particle of radius $R$ is then given by $\eta =
\rho \eta_{max}$ with the ratio of spherical particle radii being
$\rho = R / R_{\mathrm{max}}$ and $\eta_{\mathrm{max}}= 6\pi \nu
R_{\mathrm{max}}$.

Summarizing, the Langevin
equation (\ref{eq:dle}) reads in dimensionless variables:
\begin{equation}
  \label{eq:le}
  \frac{\mathrm{d} \vec{r}'}{\mathrm{d} t'} = \frac{f}{\rho} \, \vec{e}_{x}+
  \sqrt{\frac{1}{\rho}} \, \vec{\xi}(t')\, ,
\end{equation}
where the dimensionless force parameter \cite{reguera_prl, biosystems}
\begin{equation}
  \label{eq:force}
  f = \frac{L F}{k_{\mathrm{B}T}} \, .
\end{equation}
For the sake of better readability, we shall skip all the primes in
the following and proceed, if not mentioned explicitly otherwise,
with dimensionless variables.

The corresponding Fokker-Planck equation for the time evolution of the
probability distribution $P(\vec{r}, t)$ takes the form \cite{hanggi1982}
\begin{equation}
  \label{eq:fp}
  \frac{\partial P(\vec{r}, t)}{\partial t} = - \vec{\nabla}
  \vec{J}(\vec{r}, t)\, ,
\end{equation}
where $\vec{J}(\vec{r}, t)$ is the probability current:
\begin{equation}
  \label{eq:pc}
  \vec{J}(\vec{r}, t) = \frac{1}{\rho} \, \left( f \vec{e}_{x} -
    \vec{\nabla} \right) P(\vec{r}, t)\, .
\end{equation}

\subsection{Boundary conditions}

As the particles are confined by the channel structure, the
probability current has to vanish at the boundaries.
Due to the finite
size of the particles their center position can aproach the boundary
only up to its radius. Consequently, the position vector $\vec{r}$ of
a particle with radius $R$ never approaches the channel walls and is
restricted to only a portion of the inner channel area,
cf. Fig.~\ref{fig:tube}. The {\itshape effective boundary function}
$\omega^{\mathrm{eff}}(x)$, which serves as boundary for the center of
mass, exhibits the distance $R$ from the original, {\itshape true}
boundary function $\omega(x)$. Consequently, the ``no-flow'' boundary
conditions for the center of mass dynamics read:
\begin{equation}
  \label{eq:bc}
  \vec{J}(\vec{r}, t) \cdot \vec{n} = 0\, , \quad \mathrm{for }\quad \vec{r}
  \in\, \mathrm{effective \, boundaries ,}
\end{equation}
where $\vec{n}$ denotes the normal vector field at the effective
channel walls. For the considered 2D channel structure, the boundary
condition becomes
\begin{equation}
  \label{eq:bc2d}
  \frac{\mathrm{d} \omega^{\mathrm{eff}}(x)}{\mathrm{d} x} \left\{ f
    P(x,y,t) - \frac{\partial P(x,y,t)}{\partial x}\right\} +
  \frac{\partial P(x,y,t)}{\partial y} = 0\, ,
\end{equation}
at $y=\pm \omega^{\mathrm{eff}}(x)$.

\begin{figure}[t]
  \centering
  \includegraphics{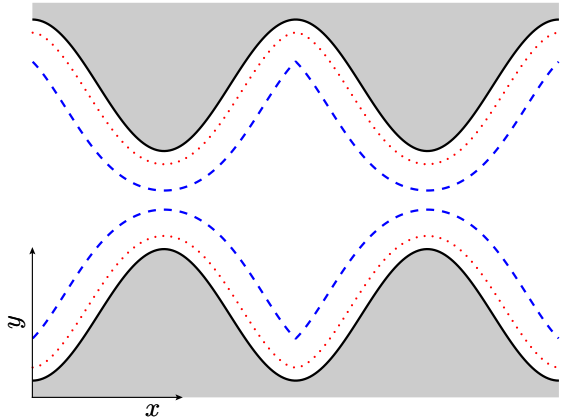}
  \caption{(Color online) Sketch of the original tube geometry given by the boundary
    function $\omega(x)$ and the effective boundary function
    $\omega^{\mathrm{eff}}(x)$ for the center of the spherical
    particle. Hereby, the effective boundary function depends on the
    radius of the particle which is given in dimensionless units by
    $\rho$: $\rho = 0.27$ (red dotted line), $\rho = 0.81$ (blue
    dashed line).}
  \label{fig:effbc}
\end{figure}

Note, that the effective boundary function exhibits a complex
dependence on the particle's radius $R$ and could not be given
explicitly. If the curvature of the channel wall function $\omega(x)$
is larger than that of the particle, the effective boundary function
exhibits a kink, cf. Fig. \ref{fig:effbc}.

For an arbitrary form of $\omega (x)$, the
boundary value problem defined by Eqs.~(\ref{eq:fp}),
(\ref{eq:pc})  and (\ref{eq:bc2d}) is very difficult to
solve. Despite the inherent complexity of this problem an
approximate solution can be found  by introducing an effective
one-dimensional description where geometric constraints and
bottlenecks are considered as {\itshape entropic barriers}
\cite{reguera_prl, burada_pre, jacobs, zwanzig, reguera_pre, percus,
  kalinay, bradely}.

\section{Fick-Jacobs approximation}

The 1D equation is obtained from the full 2D Smoluchowski equation
upon the elimination of the transversal $y$ coordinate assuming fast
equilibration in the transversal channel direction.

\subsection{The Fick-Jacobs equation}

The marginal probability density along the axis of the channel is
defined by
\begin{equation}
  \label{eq:mpd}
  {\cal P}(x,t) =
  \int_{-\omega^{\mathrm{eff}}(x)}^{\omega^{\mathrm{eff}}(x)} P(x,y,t)
  \mathrm{d}y\, .
\end{equation}
Assuming fast equilibration in $y$-direction the 2D probability
distribution becomes
\begin{equation}
  \label{eq:mpdeq}
  P(x,y,t) = {\cal P}(x,t) \, Q(y|x)\, ,
\end{equation}
with the local equilibrium distribution $Q(y|x)$ of $y$, conditional on
a given $x$. If there is no force component in transversal channel
direction (as it is in our case), the conditional distribution
$Q(y|x)$ is uniform and reads due to the normalization condition:
\begin{equation}
  \label{eq:cd}
  Q(y|x) = 1 / (2 \, \omega^{\mathrm{eff}}(x)) \, .
\end{equation}

Then on integrating the full 2D Smoluchowski equation (\ref{eq:fp})
and making use of Eqs.~(\ref{eq:mpd}), (\ref{eq:mpdeq}) and
(\ref{eq:cd}),  the Fick-Jacobs equation for the spherical particle is
obtained:
\begin{equation}
  \label{eq:fj}
  \frac{\partial \, {\cal P}(x,t)}{\partial x} = \frac{1}{\rho}
    \frac{\partial}{\partial x} D(x) \left\{ \frac{\mathrm{d}
        A(x)}{\mathrm{d} x} + \frac{\partial}{\partial x}\right\}
    {\cal P}(x,t)\, ,
\end{equation}
with the dimensionless free energy $A(x) = -f \, x - \ln
\omega^{\mathrm{eff}}(x)$. For a periodic channel this free energy
assumes the form of a tilted periodic potential with the bottlenecks
forming entropic potential barriers. Note, that for a
straight channel, i.e. constant effective boundary function, the
entropic  contribution vanishes and the particle is solely driven by
the external force.

Introducing the $x$-dependent diffusion coefficient $D(x)$ in
Eq.~\ref{eq:fj} considerably improves the accuracy of the kinetic
equation, extending its validity to more winding structures \cite{zwanzig, reguera_pre, percus,
  kalinay, bradely}. The expression for $D(x)$  (in dimensionless
units)
\begin{equation}
  \label{eq:dx}
  D(x) \doteq \frac{1}{\left[1 + \left( \mathrm{d} \omega(x) / \mathrm{d}
        x\right)^{2}\right]^{1/3}}\, ,
\end{equation}
has been shown to appropriately account for curvature effects of
the confining walls \cite{reguera_pre}.

\subsection{Nonlinear Mobility}

Besides the effective diffusion coefficient, the average particle
current, or equivalently the nonlinear mobility serves as key quantity
of particle transport through periodic channels. For any non-negative
force the average particle current in periodic structures can be
obtained from Ref.~\cite{reimann2001, lindner2001, reimann2002}
\begin{equation}
  \label{eq:apc}
  \langle \dot{x} \rangle = \langle t(x_{0} \to x_{0} +1 ) \rangle
  ^{-1}\, ,
\end{equation}
where $\langle t(a \to b) \rangle$ denotes the mean first passage time
of particles starting at $x=a$ to arrive at $x=b$. Within the
Fick-Jacobs equation (\ref{eq:fj}), the mean first passage time can be
determined:
\begin{equation}
  \label{eq:mfpt}
  \langle t(a \to b) \rangle = \rho \, \int_{a}^{b} \mathrm{d}x
  \exp(-f\, x) / \omega^{\mathrm{eff}}(x) \int_{-\infty}^{x}
  \mathrm{d}y \exp( f\,  y) \omega^{\mathrm{eff}}(y) \, .
\end{equation}

The nonlinear mobility $\mu(f)$ is defined by $\mu(f) = \langle
\dot{x} \rangle / f$  and can be obtained as
\begin{equation}
  \label{eq:nm}
  \mu(f) = \frac{1}{\rho} \cdot \frac{1-\exp(-f)}{f \displaystyle \int_{0}^{1}
    \mathrm{d}z I(z,f)} \, ,
\end{equation}
where
\begin{equation}
  \label{eq:int}
  I(z,f) = \exp(-f\, x) / \omega^{\mathrm{eff}}(x) \int_{x-1}^{x}
  \mathrm{d}y \exp( f\,  y) \omega^{\mathrm{eff}}(y) \, .
\end{equation}

In case of a straight channel with $\omega_\mathrm{min} =
\omega_{\mathrm{max}}$, an exact analytical solution of the full 2D
Smoluchowski equation (\ref{eq:fp}) is known and the nonlinear
mobility equals the free  mobility (i.e. without geometrical
constrictions)
\begin{equation}
  \mu = \mu_{\mathrm{free}} = 1 / \rho = R_\mathrm{max}/R  \quad
  \mathrm{(for \, straight\, channels)}\, .
  \label{eq:mfree}
\end{equation}
Consequently, the influence of the confinement can be expressed by
the ratio of nonlinear mobility for the transport through the channel
and the one for the unrestricted case:
\begin{equation}
  \label{eq:nmgi}
  \frac{\mu(f)}{\mu_{\mathrm{free}}} =  \frac{1-\exp(-f)}{f
    \displaystyle \int_{0}^{1}
    \mathrm{d}z I(z,f)} \, .
\end{equation}

\section{Precise numerics for a two-dimensional channel geometry}

The nonlinear mobility, predicted analytically within the Fick-Jacobs
approximation, has been compared with Brownian dynamic simulations
performed by a numerical integration of the full 2D Langevin equation
(\ref{eq:le}), using the stochastic Euler algorithm. As
random number generator we used the Box-Muller- and MT19937-algorithm
from the {\itshape GSL} library. The sinusoidal shape of the
considered two-dimensional channel is described by
\begin{equation}
  \label{eq:bf}
  \omega(x) := a \sin(2 \pi x) + b\, ,
\end{equation}
with the two dimensionless channel parameters $a$ and $b$. In physical
units, these two parameters are given by $a L$ and $b L$,
respectively. Note that $\omega(x)$ may
also be regarded as the first terms of the Fourier series of a more complex
boundary function. Due to the symmetry with respect to the $x$-axis,
the boundary function could be given in terms of the maximum
half-width $\omega_{\mathrm{max}}=b + a$ and the aspect ratio of
minimum and maximum channel width $\epsilon = \omega_{\mathrm{min}} /
\omega_{\mathrm{max}}$ (with $\omega_{\mathrm{min}}=b-a$), i.e.
\begin{eqnarray}
  \label{eq:bfrs2}
  \omega(x) &=& \frac{\omega_{\mathrm{max}} - \omega_{\mathrm{min}}}{2}\, \left[ \sin\left( 2\pi x \right) +
    \frac{\omega_{\mathrm{max}} + \omega_{\mathrm{min}}}{2} \right]\, ,\\
  \label{eq:bfrs}
  \omega(x) &=& \frac{\omega_{\mathrm{max}}}{2}\,
    \left(1-\epsilon\right)\, \left[ \sin\left( 2\pi x \right) +
      \frac{1+\epsilon}{1-\epsilon} \right]\, .
\end{eqnarray}
To ensure, that the spherical particles of radius $\rho$ stay within
this channel geometry, the integration was carried out performing ``no-flow'' boundary
conditions at the channel walls. By averaging over $10^{5}$
simulations we obtain the steady-state average particle current
\begin{equation}
  \label{eq:apcn}
  \langle \dot{x} \rangle = \lim_{t\to \infty}\, \frac{\langle x(t)
    \rangle}{t}\, ,
\end{equation}
and the nonlinear mobility $\mu =  \langle \dot{x} \rangle / f$.

\subsection{Nonlinear mobility: force and temperature - dependence}

\begin{figure}[t]
  \centering
  \includegraphics{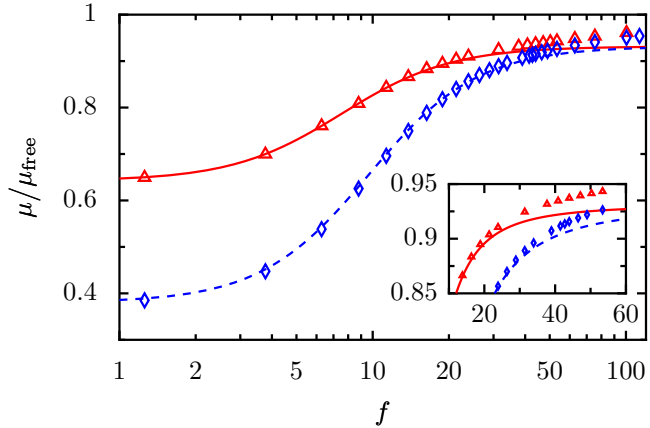}
  \caption{(Color online) Graph for the scaled nonlinear mobility as function of the
    force parameter $f$. In the Langevin simulation the different symbols correspond to different particle
    radius of $\rho = 0.2$ (red triangles) and $\rho = 0.8 $ (blue
    diamonds). The relative error of the simulation results is
    smaller than $0.01$. The Fick-Jacobs results, Eq.~(\ref{eq:nmgi}),
    correspond to the solid lines. The boundary function reads:
    $\omega(x) =  (0.7/2\pi) \sin( 2 \pi x) + 1.02/2\pi$. }
  \label{fig:nmcan}
\end{figure}

Figure \ref{fig:nmcan} depicts the nonlinear mobility as a function
of the scaling parameter $f$ for two different particle radii and a
fixed channel geometry: $\omega(x) =  0.7/(2\pi) \sin( 2 \pi x) +
1.02/(2 \pi)$. Strikingly, the transport through such channel
structures is distinctly different from the one occurring in
one-dimensional periodic {\itshape energetic} potentials
\cite{reguera_prl, biosystems, burada_pre,  burada_ptrsa}. This
phenomenon is due to the different temperature dependence of the
barrier shapes. Decreasing the temperature in an energetic periodic
potential decreases the transition rates from one cell to the
neighboring one by decreasing the Arrhenius factor \cite{hanggi1990}
and, therefore, reduces the nonlinear mobility. For the periodic
channel system, a decrease of temperature results in an increase of
the dimensionless force parameter $f$, cf. Eq.~(\ref{eq:force}) and
consequently, in a monotonic increase of the nonlinear mobility.

Due the geometrical restrictions, the nonlinear mobility is always
smaller than the mobility for the free case,
cf. Fig.~\ref{fig:nmcan}. With increasing scaling parameter,
the nonlinear mobility tends to that of the free case, i.e. $\mu \to
\mu_{\mathrm{free}}$ for $f\to \infty$.

A comparison of the analytics obtained by means of the
Fick-Jacobs-approximation with the precise numerics enables one to
determine validity criteria for the Fick-Jacobs approximation, for
further details see Ref. \cite{biosystems, burada_pre}. According to
them the applicability of the Fick-Jacobs approximation for the
transport of point particles depends on the smoothness of the geometry
and the scaling parameter $f$. For finite size particles, the diameter
of the particles should become an additional parameter in the
validity criteria. In particular, the particle's radius determines the
maximum width of the effective channel structure. A larger particle
leads to a smaller maximum effective width. Since for a fast
equilibration in the orthogonal tube direction the timescale for the
orthogonal diffusion process must be smaller than the timescale for
the drift \cite{biosystems}, a smaller maximum channel width favors the
validity of the Fick-Jacobs approximation. Thus, with increasing
particle size the range of the applicability of FJ increases,
cf. Fig.~\ref{fig:nmcan}.

However, it turned out, that the transport phenomena presented
below, occur for channel structures for which the validity criteria is
not fulfilled. Therefore, we stick in the following, to the numerical
results only.

\subsection{Particle size}

Surely, the transport of particles through small channel systems
depends on the size of the particles. In particular, the effect of
the size on the nonlinear mobility is two-fold. Firstly, the
friction coefficient depends on the particle size, resulting in a
$\rho$-dependence of the nonlinear mobility even for the case of
unconstrained motion (free case), cf. Eq.~(\ref{eq:mfree}). With
increasing radii ratio $\rho$, the mobility declines. Secondly, there is
the influence of the geometrical confinement. The extent of the
bottleneck ($2 \omega_\mathrm{wmin}$) gives a limit to the size of
particles able to travel through the channel. In our scaling the
largest sphere possible to overcome the geometry's bottleneck has a
radii ratio of $\rho = 1$. Considering particles of different sizes,
the effective bottleneck ($2 \omega_\mathrm{min}^\mathrm{eff}$) will
be smaller for spheres of higher diameter. It is intuitive that a
small bottleneck hinders the transport.

\begin{figure}[t]
  \centering
  \includegraphics{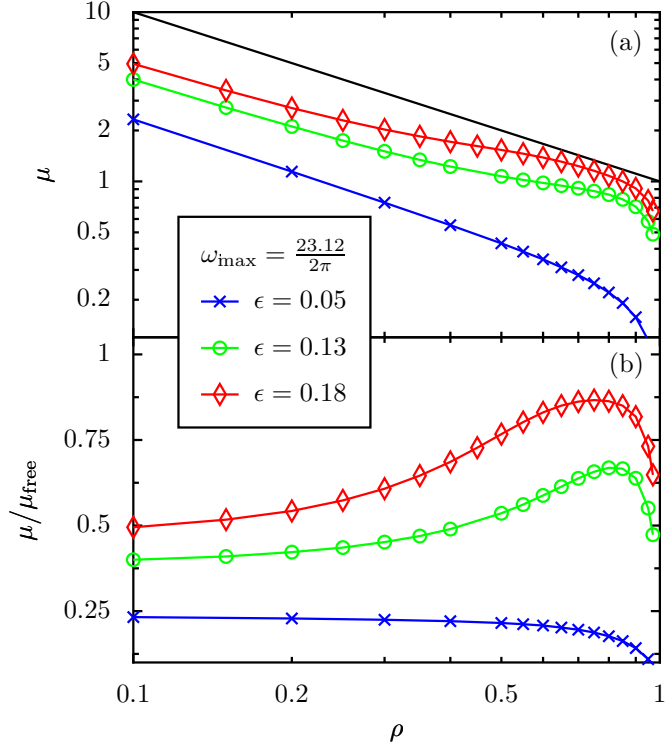}
  \caption{(Color online) The numerically obtained non-linear mobility $\mu$ (a) and
    the scaled nonlinear mobility $\mu/\mu_{\mathrm{free}}$ (b) as function of
    the radii ratio $\rho = R / R_{\mathrm{max}}$ for different channel geometries
    ({\itshape constant-width-scaling}: $\omega_{\mathrm{max}}=\mathrm{const}$), cf. Eq.
    (\ref{eq:bfrs}). }
  \label{fig:nm}
\end{figure}

Figure~\ref{fig:nm}(a) depicts the nonlinear mobility as function of
the radii ratio $\rho$ for different geometries. For a straight
channel one observes the nonlinear mobility of the free case
$\mu_{\mathrm{free}}$ which depends reciprocally on the radii ratio
$\rho$. In presence of geometrical restrictions, i.e. for varying
cross-section width, the nonlinear mobility is smaller than
$\mu_{\mathrm{free}}$, cf. Fig.~\ref{fig:nm} (a). Moreover, upon
decreasing the bottleneck half-width (i.e. decreasing the aspect
ratio $\epsilon$) of the structure, the nonlinear mobility decreases
\cite{burada_appb}.

Deviations from the $1/\rho$-dependence show another effect of the
geometrical confinement, cf. Fig.~\ref{fig:nm} (a).
In order to focus on the geometrical effect, we consider
the scaled nonlinear mobility, i.e. the nonlinear mobility relative to
the nonlinear mobility in free case: $\mu / \mu_{\mathrm{free}}$,
cf. Fig.~\ref{fig:nm} (b). This is equivalent to the consideration of the
nonlinear mobility of a point particle moving in the effective channel
geometry defined by the effective boundary function
$\omega^{\mathrm{eff}}(x)$, which still depends on the parameter
$\rho$. With increasing $\rho$, the maximum half-width of the effective
geometry shrinks. As a consequence, the sojourn time, the particle
spends on average in a bulge of the channel structure decreases with
increasing $\rho$ and the mobility of the point particle in the
effective geometry grows. This behavior causes
the maximum in the scaled nonlinear mobility and the shoulder in the
dependence of the nonlinear mobility on the radius,
cf. Fig.~\ref{fig:nm}.

\subsection{Role of the channel structure}

The confinement by the considered channel geometry can be altered by
systematically changing the parameters $\omega_{\mathrm{max}}$ and
$\omega_{\mathrm{min}}$ or $\omega_{\mathrm{max}}$ and $\epsilon$ in
the boundary function, Eq.~(\ref{eq:bfrs2}) or Eq.~(\ref{eq:bfrs})
respectively. While for Fig.~\ref{fig:nm} we examined a constant
maximum half-width $\omega_{\mathrm{max}}$ = const and varied the
aspect ratio $\epsilon$ which is equivalent to vary the half-width
$\omega_{\mathrm{min}}$ at the bottleneck, cf. the {\itshape
constant-width-scaling} in Ref.~\cite{burada_appb}, it is
instructive to also consider both, the {\itshape
constant-bottleneck-scaling} $\omega_{\mathrm{min}}$ = const as well
as the {\itshape constant-ratio-scaling} $\epsilon =
\omega_{\mathrm{min}}/\omega_{\mathrm{max}}$ = const.

\begin{figure}[t]
  \centering
  \includegraphics{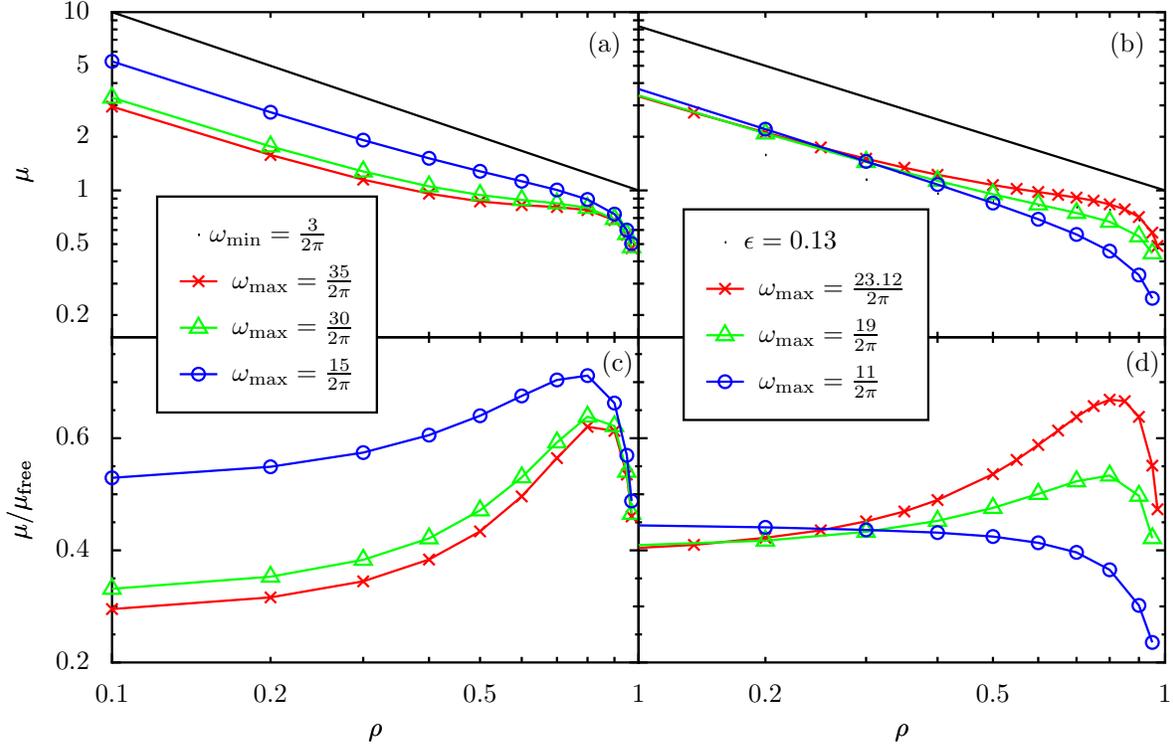}
  \caption{(Color online) The nonlinear mobility (a), (b) and scaled nonlinear
    mobility (c),(d) are depicted for different scalings of the
    geometry: constant-bottleneck-scaling,
    i.e. $\omega_{\mathrm{min}}$ = const, in (a) and (c) ;
    constant-ratio-scaling, i.e. $\epsilon =\omega_{\mathrm{min}} /
    \omega_{\mathrm{max}}$ = const, in (b) and (d).  }
  \label{fig:nm2}
\end{figure}

As we keep the bottleneck-width constant and decrease the maximum
half-width, the sojourn times the particles spends in the bulges
decreases causing the mobility to increase and approach the maximum
value for a straight channel
($\omega_{\mathrm{min}}=\omega_{\mathrm{max}}$ and $\epsilon=1$),
cf. Fig.~\ref{fig:nm2} (a).  In contrast, within the {\itshape
constant-ratio-scaling}, where the bottleneck half-width
 scales with the maximum half-width, the nonlinear
mobility $\mu$ depends for small radii only slightly on
$\omega_{\mathrm{max}}$, cf. Fig.~\ref{fig:nm2} (b). However, with
increasing particle size, i.e. radii ratio $\rho$, the
nonlinear mobility $\mu$ shows a striking dependence on the maximum
half-width.

As pointed out already, with increasing particle radius  the
effective maximum half-width decreases. The effective maximum
half-width will exhibit a linear dependence on the radius if the particles
curvature is larger than that of the channel's boundary function (in
physical units),
\begin{equation}
  \label{eq:curve}
  \omega_{\mathrm{max}}^{\mathrm{eff}} = \omega_{\mathrm{max}} -
  R\, , \quad \mathrm{for }\quad 1/R > \frac{- \mathrm{d}^{2}
    \omega(x_{\mathrm{max}}) / \mathrm{d}x^{2}}{
    \left[ 1 + \left( \mathrm{d} \omega(x_{\mathrm{max}}) / \mathrm{d}
        x\right)^{2}\right]^{3/2}}\, ,
\end{equation}
where $x_{\mathrm{max}}$ denotes the $x$-values for which the boundary
function assumes a maximum. For larger particle radii the effective
boundary function shows a kink and the effective maximum half-width
$\omega^{\mathrm{eff}}_{\mathrm{max}}$ decreases faster than linearly
with the radii ratio $\rho$. As the sojourn times the particle spend in the
channel's bulges depends mainly on the effective maximum width, a
nonlinear dependence of the nonlinear mobility is observed for larger
particle radii causing a peak in the scaled nonlinear
mobility, cf. Fig.~\ref{fig:nm2} (c) and (d).

\section{Conclusions and Outlook}

We studied the transport of finite Brownian particles through
channels with periodically varying width. For point size particles
it was shown previously \cite{biosystems, burada_pre}, that the
transport through such channels could be approximately described by
means of the so-called Fick-Jacobs equation which is based on the
assumption  of a  fast equilibration in orthogonal transport
direction. Validity criteria for the capability of this
approximation include a dependence on the channel shape and predict
an upper limit for the force value. In case of spherical, finite
size particles the maximum force value up to which the Fick-Jacobs
equation could be applied depends also on the size of the particle.
By comparison of the approximative result for the nonlinear mobility
and the numerical ones we have shown, that the equilibration
assumptions holds for a wider force-range in case of larger
particles than it is the case for smaller ones.

In addition, we pointed out, that the transport of finite,  spherical Brownian
particles in channel geometries with highly corrugated channel walls
exhibits some striking features which may allow for the development
of newly separation devices which extends the functionality of the
sieves. In particular we found, that the nonlinear mobility of
Brownian particles in such channel structures deviates from the
one-over-size dependence predicted by the Stokes law for Brownian
particles moving in an environment without geometrical
constrictions. Instead, there is an optimal particle size for which
the nonlinear mobility as compared to the free mobility exhibits a
maximum value.

Our present study also implicitly used a  small concentration of spherical particles such that both, effects of
particle-particle  interactions and  forces between particle-particle and particle-walls due to hydrodynamic interactions
can safely be ignored. These complications would require totally new and extensive studies that are beyond this present study.
Moreover, as emphasized with the abstract already, we assumed throughout
perfect spherical symmetry.
Deviations from such spherical symmetry would also impact the viscous friction law
behavior \cite{koenig} and, as well, may give rise to additional, new entropic effects. All such complications
are beyond the work presented here; all these latter complications, however,
open up avenues for interesting future investigations.

\section*{Acknowledgments}
This work has been supported by the Volkswagen foundation (project
I/83 902), the Max-Planck society and by the German Excellence Initiative via the Nanosystems
Initiative Munich (NIM).

\end{document}